\documentclass{iopconfser}

\usepackage[]{xcolor}
\usepackage[most]{tcolorbox}
\usepackage{amsmath}

\DeclareMathOperator{\cn}{cn}

\usepackage[utf8]{inputenc}
\usepackage{csquotes}
\usepackage[backend=bibtex,doi=false,isbn=false,url=false,eprint=false]{biblatex}
\usepackage{amssymb}
\usepackage{txfonts}
\usepackage[english]{babel}
\usepackage{mathtools}
\usepackage{physics}
\usepackage{multicol}
\usepackage{graphicx}
\usepackage{wrapfig}
\usepackage[margin=5pt,font=small,labelfont=bf,labelsep=endash]{caption}
\usepackage{placeins}
\usepackage{slashed}
\usepackage{lipsum}
\usepackage{feynmf}
\usepackage{array}
\usepackage[thinlines]{easytable}
\usepackage{makecell}

\DeclareFieldFormat[article]{volume}{\textbf{#1}}
\renewbibmacro{in:}{}

\DeclareBibliographyDriver{article}{%
  \printnames{author} 
\space 
  \printfield{year} 
\space 
  \printfield{title} 
\space 
  \printfield{journaltitle} 
\space 
  \printfield{volume} 
\printfield{number}
  \printfield{pages} 
}

\begin{document}

\title{Higher Time Derivative Theories From Integrable Models}

\author{Bethan Turner$^1$}

\affil{$^1$Department of Mathematics, City, University of London, London, UK}
\email{bethan.turner.2@city.ac.uk}

\begin{abstract}
Higher Time Derivative Theories are generated by considering space-time roated KdV and mKdV systems. These systems are then studied to see if/how instabilities, usually associated with higher time derivative theories, manifest on the classical level by presenting both analytic and numerical solutions. For a linearised version of these space-time rotated systems we present a detailed quantisation of the theory that highlights the known dilemma on higher time derivative theories, that we have either negative norm states or the Hamiltonian being unbounded from below.
\end{abstract}

\section{Introduction}
\label{intro}
 Higher Time Derivtive Theories (HTDT) are generally discounted as unphysical on the grounds that they contain fundamental instabillities, often referred  to in the literature as ghosts. These arise since for any HTDT the higher dimension of the configuration space necessitates a higher dimensional phase space and this inevitably leads us to a Hamiltonian that is linear in all phase space momenta bar one, and is thus unbounded from below. This has been known for a long while with the first work on this (at the classical level) being carried out by Ostrogradsky in \cite{Ostrogradsky1850} (see \cite{Orstogradskytheorem} for a more recent English language review). In the absence of interaction terms, this will still allow for stable dynamics, however if we allow interaction terms then there is nothing restricting how much energy can be exchanged between positive and negative energy modes, and this is generally expected to lead to unstable dynamics that then very often run away to infinity. Subsequent study of quantum HTDT's has found that at the quantum level a bounded Hamiltonian can be achieved but only at the cost of unitarity \cite{paisuhlenbeck}. However, such theories do possess a number of interesting properties, chief among them that they tend to be renormalisable, and therefore can, for example, be used to construct renormalisable theories of gravity \cite{stelle,Starobinsky1980}. They also have other applications for instance in finite temperature problems \cite{Weldon_1998} and the study of black holes \cite{blackholes}. 

Given the possible applications of HTDT's attempts are often made to resolve or get around the issue of the instabilities, see for example \cite{Hawking_2002,Exorcism}. Indeed it is well known that despite these instabilities some HTDTs can admit stable dynamics, even in the presence of interaction terms \cite{Pagani1987,Salvio_2019}.  It has therefore been suggested that we can distinguish between malicious and benign ghosts, \cite{Smilga_2005}. In this nomenclature, (classical) systems with benign ghosts admit dynamics where the trajectory goes too infinity in infinite time, whereas if the ghosts are malicious, trajectories go to infinity in finite time. Even if we have stable analytic solutions, a system can still be said to contain malicious ghosts if its dynamics diverge under perturbation, if they remain stable they are benign.  

 It has also been suggested that integrability may play some role in determining which kind of ghost we encounter \cite{deffayet2022}. For example in \cite{Toda,Calogero} we found that in systems with ghost ridden Hamiltonians, the resulting dynamics were stable precisely because they were integrable. Here we present some results of a study based on an idea put forward in \cite{Smilga:2021nnx,DSBG} and consider space-time rotated KdV systems. While this does not change the analytic solutions beyond the need to switch $x$ and $t$, this nevertheless changes the dynamics of the system as the Cauchy problem has become one with three pieces of initial value data as opposed to one. These space-time rotated models retain the integrability of standard KdV systems, and what we have found is that the presence of higher conserved charges imposes restrictions on how a non analytic initial pulse can evolve, and seems to  influene the way in which the ghosts manifest. For a more detailed discussion of the results presented here see \cite{KDV,stability}.

\section{Hamiltonian Formalism for Systems of Rotated KdV Type}

We begin by comparing the Hamiltonian formalism for the standard and rotated KdV systems. It is well known that the standard KdV equations can be derived from the Lagrangian (see e.g. \cite{MGKIV})

\begin{subequations}
 \begin{minipage}{0.48\linewidth}
  \begin{equation}
\mathcal{L} = \frac{1}{2}\phi_t\phi_x+\phi_x^n-\frac{1}{2}\phi_{2x}^2,
\end{equation}   
 \end{minipage}
  \begin{minipage}{0.48\linewidth}
  \begin{equation}
u_t+n\left(n-1\right)u^{n-2}u_x+u_{3x}=0, 
\label{KdVEOM}
\end{equation}   
 \end{minipage}
\label{KdVLEOM}
 \end{subequations}
where $u=\phi_x$.  The cases of $n=3$ and $n=4$ correspond to well known integrable systems related by the Mirua transform \cite{KGM1,KGM2}. We are mostly interested in these systems, although towards the end we will quantise a model where $n=2$. The phase space variables can be found in the usual way, from which the standard KdV poisson bracket and Hamiltonian can be derived, 

\begin{subequations}
\begin{minipage}{0.14\linewidth}
\begin{align}
\Phi=& \phi, 
\end{align} 
\end{minipage}
\begin{minipage}{0.15\linewidth}
\begin{align}
\Pi=& \frac{1}{2}\phi_x, 
\end{align} 
\end{minipage}
\begin{minipage}{0.35\linewidth}
\begin{align}
\{u_x\left(x\right),u\left(x'\right)\}= & -\delta_x\left(x-x'\right),
  \label{poissbrackkdv}
\end{align}
\end{minipage}
\begin{minipage}{0.32\linewidth}
\begin{align}
\mathcal{H}=&\frac{1}{2}u_x^2-n\left(n-1\right)u^3,
\end{align}
\end{minipage}
\end{subequations}
where $\mathcal{H}$ generates time evolution in the usual way \footnote{Technically this system is degenerate as $\Pi$ has no relation to $\phi_t$. This can be corrected via the use of Dirac constraints, see e.g. \cite{Nutku}. The Dirac constraints are not needed to recover the eqution of motion however.} \cite{Faddeev1985}. We now introduce higher time derivatives by interchanging $x$ and $t$ in equation (\ref{KdVLEOM})

\begin{subequations}
\begin{minipage}{0.48\linewidth}
\begin{equation}
\mathcal{L} = \frac{1}{2}\phi_t\phi_x+\phi_t^n-\frac{1}{2}\phi_{2t}^2 
\label{Lagrangian}
\end{equation}
\end{minipage}
\begin{minipage}{0.48\linewidth}
\begin{equation}
u_x+n\left(n-1\right)u^{n-2}u_t+u_{3t}=0,
\end{equation}    
\end{minipage}
\label{rKdVLEOM}
\end{subequations}
where now $u=\phi_t$. Thus far all we have done is swap $x$ and $t$, however we have not changed the roles played by these variables, $t$ is still the evolution parameter. Thus Ostrogradsky's construction brings us to a now four dimensional phase space

\begin{subequations}
\begin{minipage}{0.19\linewidth}
\begin{align}
\Phi_1=&\phi, 
\end{align} 
\end{minipage}
\begin{minipage}{0.19\linewidth}
\begin{align}
\Phi_2=&\phi_t,
\end{align} 
\end{minipage}
\begin{minipage}{0.39\linewidth}
\begin{align}
\Pi_1=&\frac{1}{2}\phi_x+\phi_{3t}+n\phi_t^{n-1}, 
\end{align} 
\end{minipage}
\begin{minipage}{0.19\linewidth}
\begin{align}
\Pi_2=&-\phi_{2t}.
\end{align} 
\end{minipage}
\label{rKdVphasespace}
\end{subequations}
These variables give us the following non-vanishing Poisson Bracket relations,

\begin{subequations}
\begin{minipage}{0.48\linewidth}
\begin{align}
\{\phi\left(x\right),\phi_{3t}\left(x'\right)\}=&\delta\left(x-x'\right),  \\
\{\phi_{3t}\left(x\right),\phi_{3t}\left(x'\right)\}=&-\delta_x\left(x-x'\right), 
\end{align}
\end{minipage}
\begin{minipage}{0.48\linewidth}
\begin{align}
\{\phi_t\left(x\right),\phi_{2t}\left(x'\right)\}=&-\delta\left(x-x'\right), \\
\{\phi_{3t}\left(x\right),\phi_{2t}\left(x'\right)\}=&n\left(n-1\right)\phi_t^{n-2}\left(x\right)\delta\left(x-x'\right).
\end{align}
\end{minipage}
\label{rKdVPoissBrackphi}
\end{subequations}
The Hamiltonian is now

\begin{equation}
\mathcal{H}_r=\frac{1}{2}\Pi_2^2+\Pi_1\Phi_2-\frac{1}{2}\left(\Phi_1\right)_x\Phi_2-\left(\Phi_2\right)^3=uu_{2t}+\frac{6}{n}u^n-\frac{1}{2}u_t^2, 
\label{rKdVHamiltonian}
\end{equation}
which given the Poisson bracket relations in (\ref{rKdVPoissBrackphi}) generates time evolution. Note also that unlike the Lagrangian and equation of motion, this is not just the standard KdV Hamiltonian with space and time interchanged. This is because despite the fact that the the Euler-Lagrange equation is unchanged under exchange of $x$ and $t$, the same does not hold for Hamilton's equations.
Note that in the first expression (\ref{rKdVHamiltonian}) is linear in $\Pi_1$, and thus is unbounded from below in the usual way of HTDT's.

\section{Stability of Rotated KdV Solutions}

A large class of solutions for the standard unrotated KdV systems are known in explicit analytic form. There are the well known soliton solutions, which can also be viewed as the $m \to 1$ limit of broader class of Jacobi Elliptic solutions with different boundary conditions. Solutions can be combined by means of auto-B\"acklund transformations. Of course all of these solutions will also provide a solution of the unrotated equation if we simply exchange $x$ and $t$. However, rotating the problem changes the setting in two ways. First  we change both the dispersion relation, which goes from $\omega \propto \kappa^3$, in the unrotated case to $\omega \propto \kappa^{\frac{1}{3}}$ in the rotated. Second, as discussed in section \ref{intro}, we have changed the Cauchy problem to one with three pecies of initial data. It is this rotated Cauchy problem and the stability of its solutions we are interested in.

\begin{figure}[h]
\centering
\noindent
\includegraphics[scale=0.32]{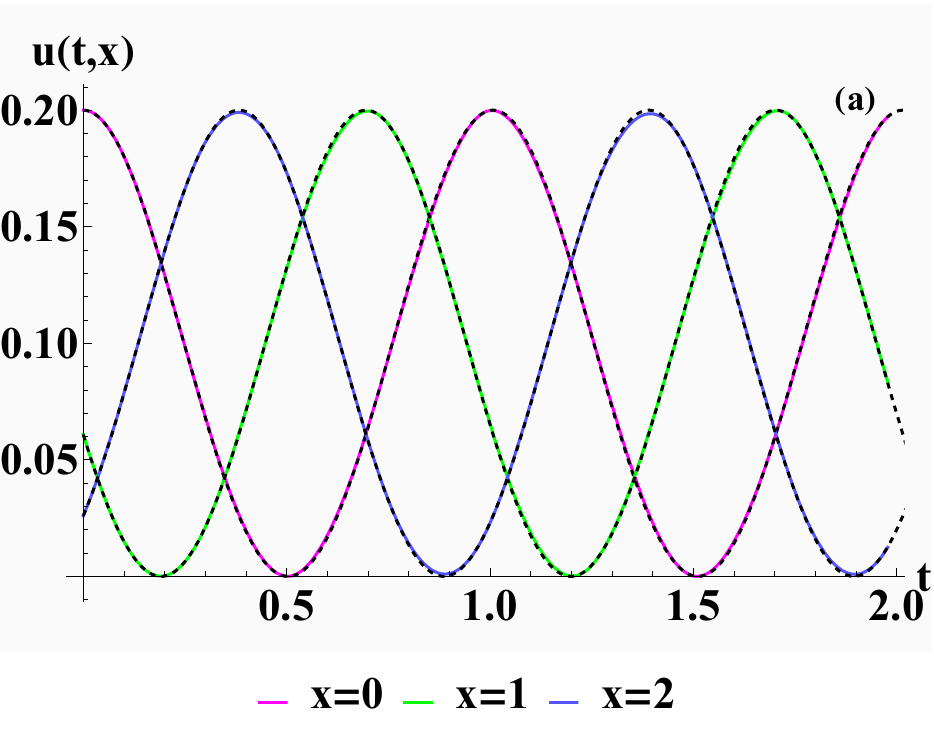}
\includegraphics[scale=0.32]{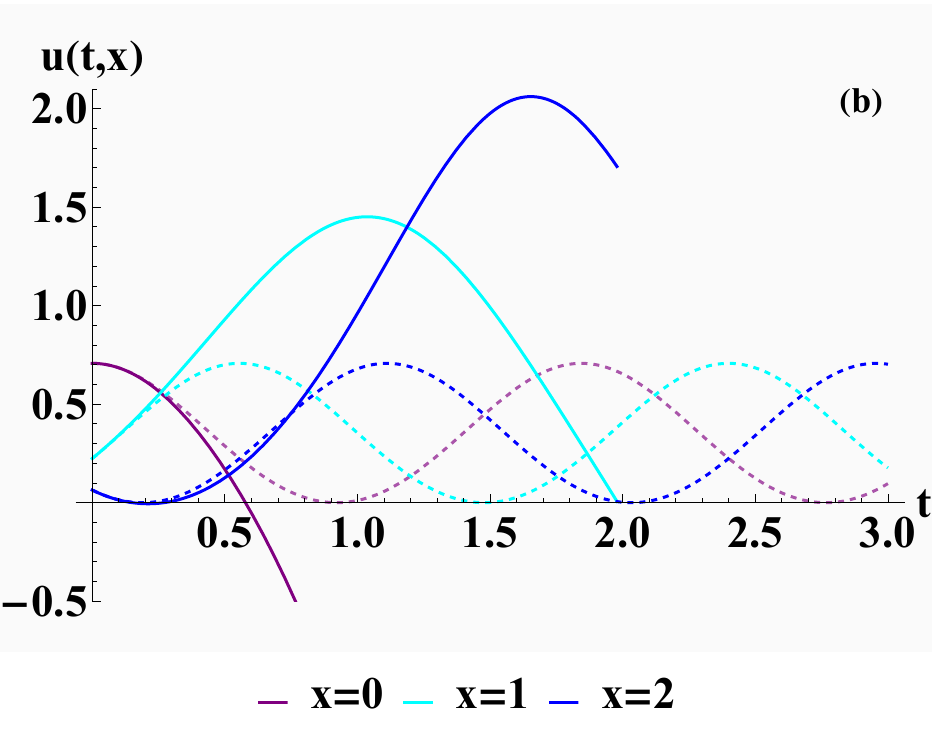} 
\includegraphics[scale=0.32]{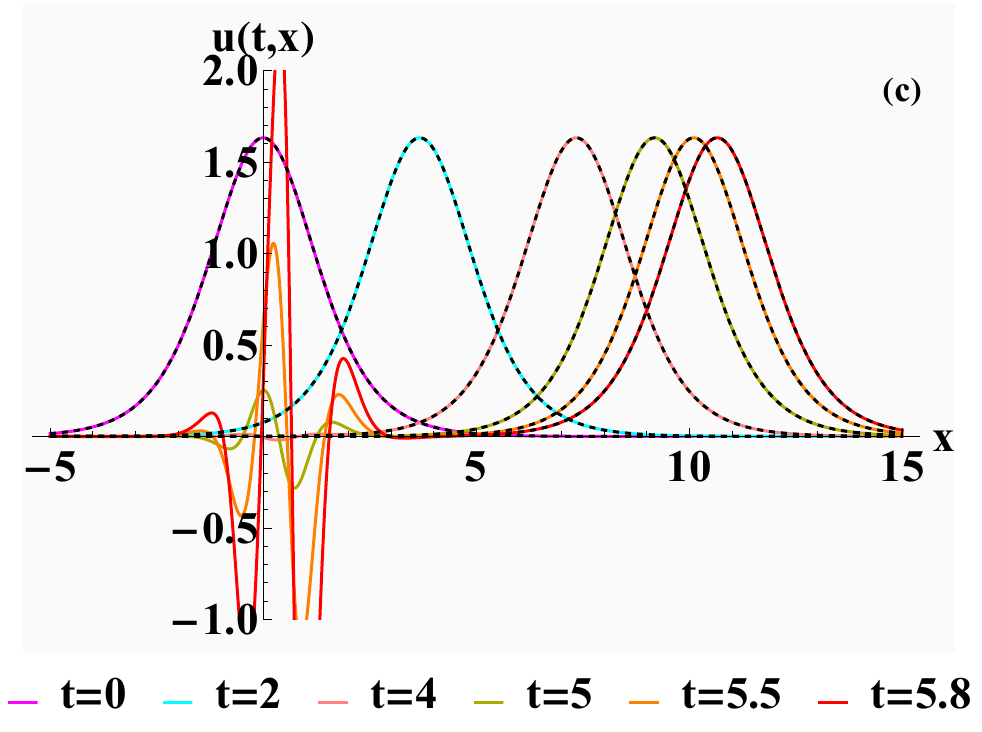}
\caption{The analytic solutions are compared with the numerical evolution of the exact initial conditions given by these solutions. In panel (a) this is done for the standard KdV equation where the solution is given by $u\left(t,x\right)=2 \kappa ^2 m \text{cn}\left(\left.4 (1-2 m) t \kappa ^3+x \kappa \right|m\right)$ (black dashed lines)  and there is one piece of Cauchy data $u\left(0,x\right)$. Panel (b) shows the results for a  space-time rotated KdV system. The analytic solution is now $u\left(t,x\right)=\frac{\kappa}{\sqrt{2}}  \text{cn}\left(\left.(m-2) t \kappa ^3+x \kappa \right|m\right)^2$ and we have three pieces of Cauchy data $u\left(0,x\right)$, $u_t\left(0,x\right)$, $u_{2t}\left(0,x\right)$, In panel (c) this is done for a soliton solution of the rotated KdV equation where $u\left(t,x\right)=2 \times 2^{1/3}\kappa^{2/3}\sech^2\left[\kappa x -\frac{\kappa^{1/3}}{4 \times 2 ^{1/3}}t\right]  $. In all cases the exact solutions are given by dashed lines and the solid lines give the numerical solutions.}
\label{kdvanvnum}
\end{figure}

A useful test of stability is to compare the numerical evolution of exact solutions, as the numerics act as a sort of pertubation of the system. In figure \ref{kdvanvnum} we present this for the elliptical solution of both the standard and rotated KdV equation, and the rotated soliton solution. For the unrotated case the numerics follow the analytical solution very well, however in the rotated case the amplitude of the oscillation tends to infinty very quickly. In the case of the soliton solution, the numerics tracks the soliton part of the solution exactly however at the origin we have the development of a standing wave the amplitude of which rapidly goes to infinity. 

\begin{figure}[h]
\centering
\begin{minipage}{\linewidth}\centering
\includegraphics[scale=0.32]{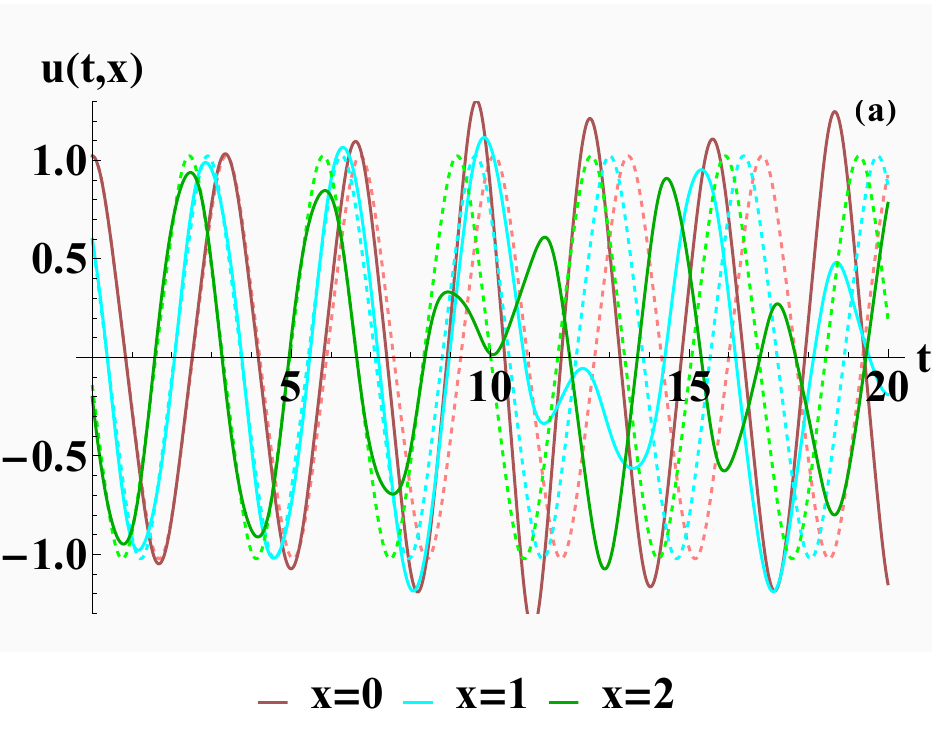}
\includegraphics[scale=0.32]{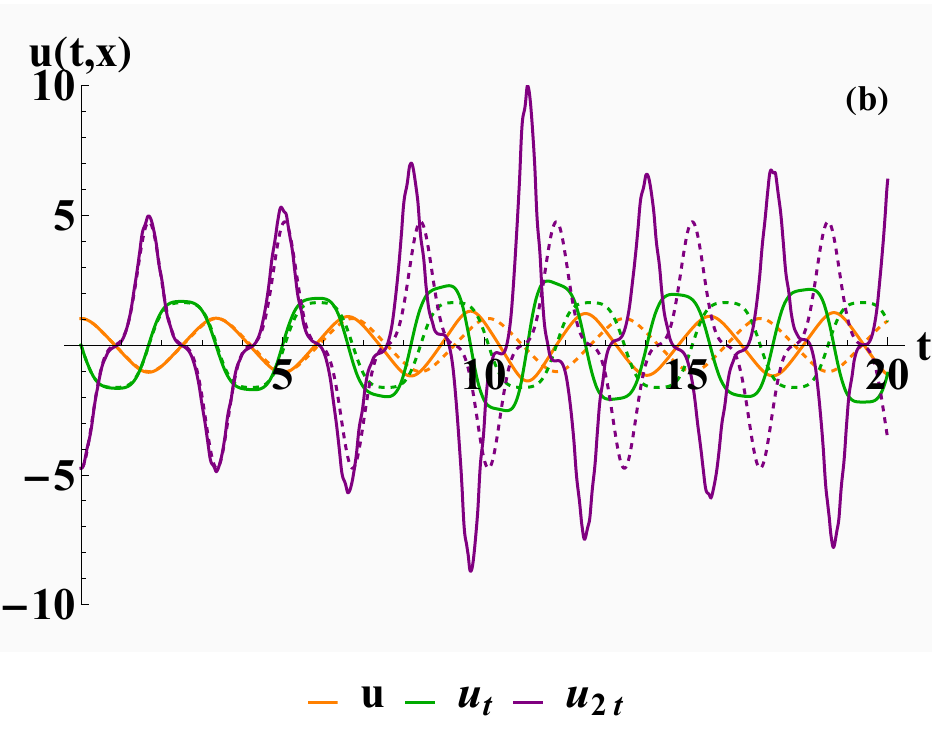}
\includegraphics[scale=0.32]{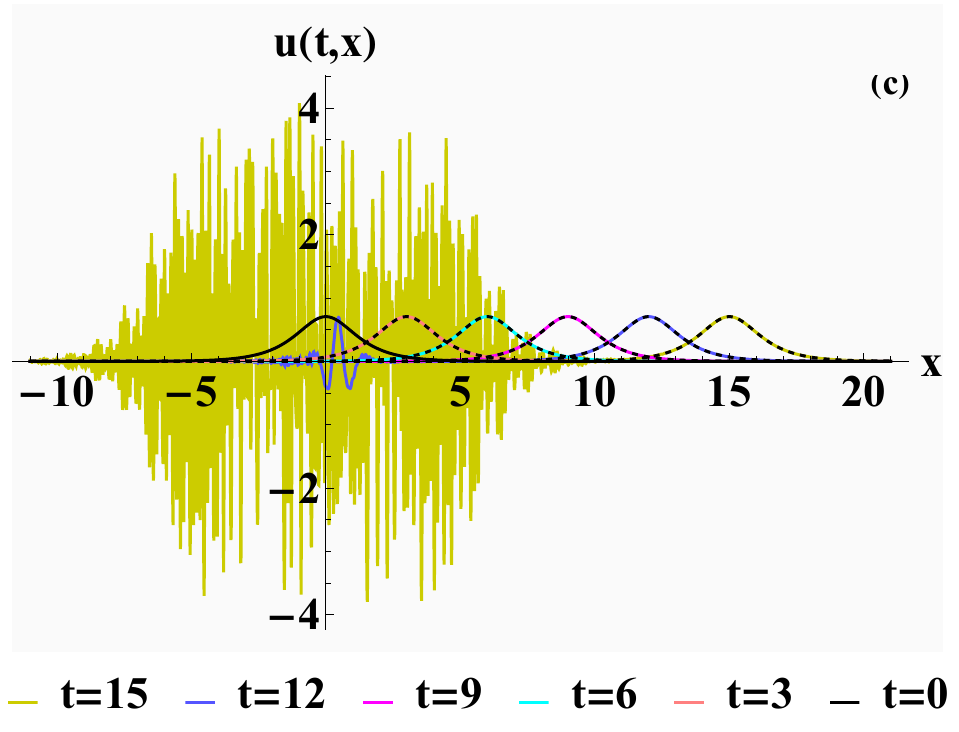} 
\end{minipage} 
 \caption{As in figure \ref{kdvanvnum} the analytic solution is compared with the numerical evolution of initial conditions, this time for the space-time rotated mKdV equation. In panel (a) the solution $u\left(t,x\right)=\sqrt{\frac{m}{2}}\left(\frac{\kappa}{1-2m}\right)^{\frac{1}{3}}\cn\left(\kappa x| m\right)$ where $\kappa=1$ and $m=0.45$ is considered. In panel (b) we plot a comparison between the Cauchy data as given by the exact solution and the numerical evolution of this data. In panel (c) this comparison between analytic and numerical evolution is done for the soliton solution $u\left(t,x\right)=-6^{2/3}\kappa^{1/3}\sech\left[\kappa x -\frac{\kappa^{1/3}}{6 ^{1/3}}t\right] $. As in figure \ref{kdvanvnum} the exact solutions are given by dashed lines and the solid lines give the numerical solutions. }
 \label{MKdVanvnum}
\end{figure}

In constrast we seem to have a different result for the rotated mKdV equation, as can be seen in figure \ref{MKdVanvnum}. Once more in the case of the ellipitcal solution is that while the numerics once more do not track the numerical solution, they also do not diverge. The same is true for the Cauchy data. In the case of the soliton solution, once more the soliton part is tracked, but we get the development of numerical noise at the origin that spreads out and eventually consumes the soliton, but it does not diverge. This indicates that while in the case of the rotated KdV equation the instability is malicous, in the case of the mKdV equation they are rather more benign. 
 
\section{Stability of Rotated KdV Type System with Gaussian Initial Pulse}
We now consider a non exact initial condition to see if this can tell us anything about why we observe this difference. We take as our starting point the argument originally put forward in \cite{Berezin1967}, see also \cite{Jeffrey1972}, that the integrability of KdV type systems governs the break up of Gaussian initial pulses into solitons. 
Now we introduce new dimensionless variables 

\begin{minipage}{0.15\linewidth}
\begin{equation}
x \to \frac{\sigma^2 \lambda_n^{3-2n}}{l}, \nonumber
\end{equation}    
\end{minipage}
\begin{minipage}{0.15\linewidth}
\begin{equation}
t \to \lambda_n t, \nonumber
\end{equation}    
\end{minipage}
\begin{minipage}{0.15\linewidth}
\begin{equation}
u \to \lambda_n u, \nonumber
\end{equation}    
\end{minipage}
\begin{minipage}{0.2\linewidth}
\begin{equation}
\lambda_n =\left[\frac{n\left(n-1\right)^3}{\sigma^2}\right]^{\frac{1}{4-3n}}, \nonumber
\end{equation}
\end{minipage}
\begin{minipage}{0.31\linewidth}
\begin{equation}
u_t+u_xu^{n-2}+\frac{1}{\sigma^2}u_{3x}=0.
\end{equation}    
\end{minipage}    

We select the Gaussian initial condition $u_0=e^{-x^2}$ and observe how it breaks up into $n$-solitons depending on the value of $\sigma$. The number of emergent solitons can be predicted by making use of the conservation laws. Assume as $t \to \infty$ that $u_0$ breaks up an $n$-soliton solution. Since the energy of $n$ asymptotically seperated solitons is equal to the sum of the energies of $n$ one soliton solutions we can write

\begin{minipage}{0.32\linewidth}
\begin{align}
Q_i \left(u_0\right)=&\sum_{i=1}^n Q_i \left(u_i\right), \nonumber
\end{align}    
\end{minipage}
\begin{minipage}{0.32\linewidth}
 \begin{align}
u_i=&a_i \sech^2\left[\frac{\sqrt{a_i}\sigma}{\sqrt{12}}\left(x-\frac{a_i}{3}t\right)\right], \nonumber
\end{align}   
\end{minipage}
\begin{minipage}{0.32\linewidth}
\begin{equation}
Q_i= \int^{\infty}_{-\infty} \mathcal{Q}_i dx.
\label{KdVConservation}
\end{equation}    
\end{minipage}
Here $\mathcal{Q}_i$ refers to the charge densities of the (m)KdV equation and $a_i$ refers to the amplitude of the $i^{th}$ soliton. We can then use (\ref{KdVConservation}) to find regions of $\sigma$ for which a Gaussian Initial Pulse will break up into $n$ solitons and predict their amplitudes. Consider, for example, the case of $n=2$, as in \cite{Berezin1967}, we can take the first two charge densities of the KdV system and generate a system of two simultaneous equations, 

\begin{subequations}
\begin{minipage}{0.14\linewidth}
\begin{equation}
\mathcal{Q}_1=u,
\end{equation}    
\end{minipage}
\begin{minipage}{0.19\linewidth}
\begin{equation}
\mathcal{Q}_2=\frac{1}{2}u^2,
\end{equation}    
\end{minipage}
\begin{minipage}{0.29\linewidth}
\begin{equation}
4\sqrt{3}\left(a_1^{\frac{1}{2}}+a_2^{\frac{1}{2}}\right)=\sqrt{\pi}\sigma,
\end{equation}    
\end{minipage}
\begin{minipage}{0.34\linewidth}
\begin{equation}
4\left(a_1^{\frac{3}{2}}+a_2^{\frac{3}{2}}\right)=\sigma\sqrt{\frac{3\pi}{8}}.
\end{equation}    
\end{minipage}
\end{subequations}
Note that in deriving these equations we have assumed $\sqrt{a}_i$ is real and positive. Upon substituting in $u_0$ and $u_i$ with $n=2$ and solving the resultant simultaneous equations we find the expressions for the amplitudes in terms of $\sigma$ shown in (\ref{ampsigmakdv}), the square roots of which are real and positive for the range of $\sigma$ given in (\ref{ampsigmakdvrange}) 
\begin{subequations}
\begin{minipage}{0.48\linewidth}
\begin{equation}
\sqrt{a_{1/2}}=\frac{1}{24}\left(\sqrt{3\pi}\sigma \pm \sqrt{144\sqrt{2}-\pi\sigma^2}\right),   
\label{ampsigmakdv}
\end{equation}
\end{minipage}
\begin{minipage}{0.48\linewidth}    
\begin{align}
6 \times 2^{1/4} \sqrt{\pi} \leq \sigma \leq 12 \times 2^{1/4} \sqrt{\pi}.
\label{ampsigmakdvrange}
\end{align}
\end{minipage}
\end{subequations}

\begin{figure}[h]
\centering
\begin{minipage}{0.67\textwidth}
\centering
\includegraphics[scale=0.5]{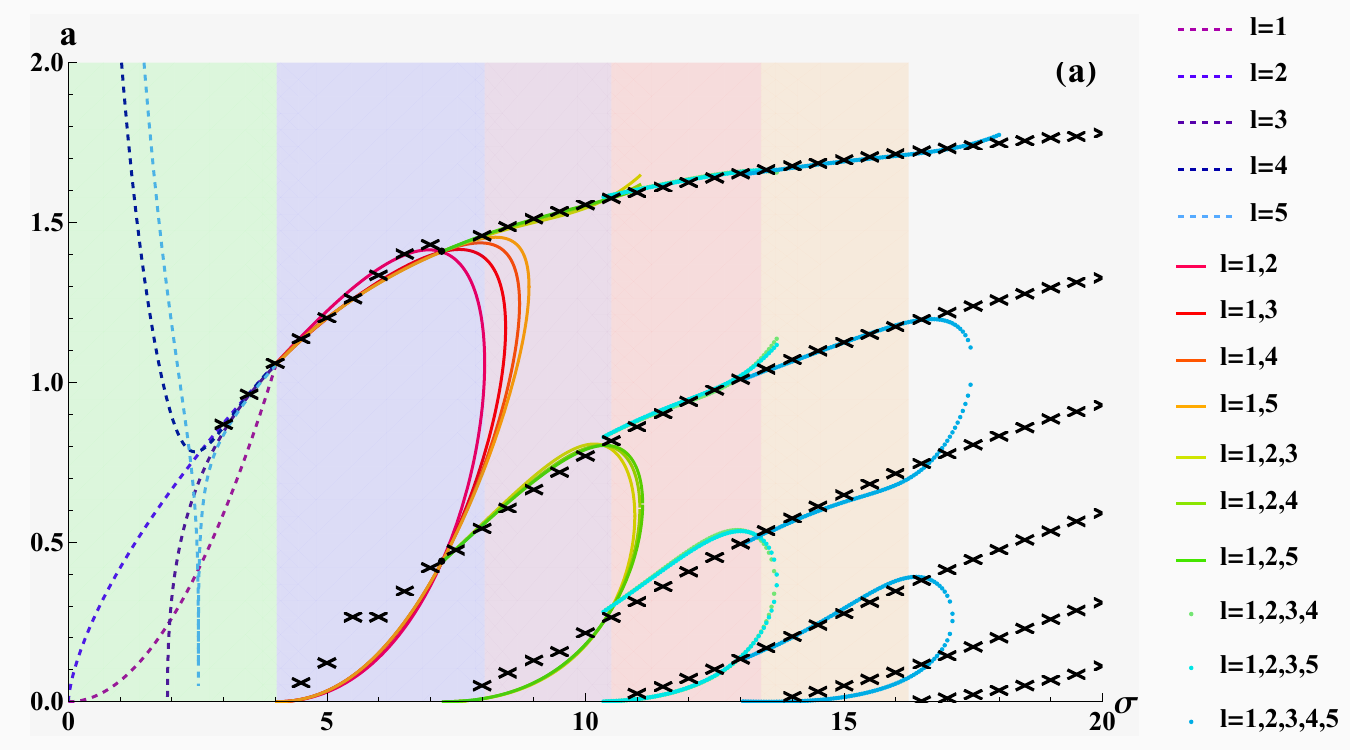}    
\end{minipage}
\begin{minipage}{0.32\textwidth}
\centering
\includegraphics[scale=0.25]{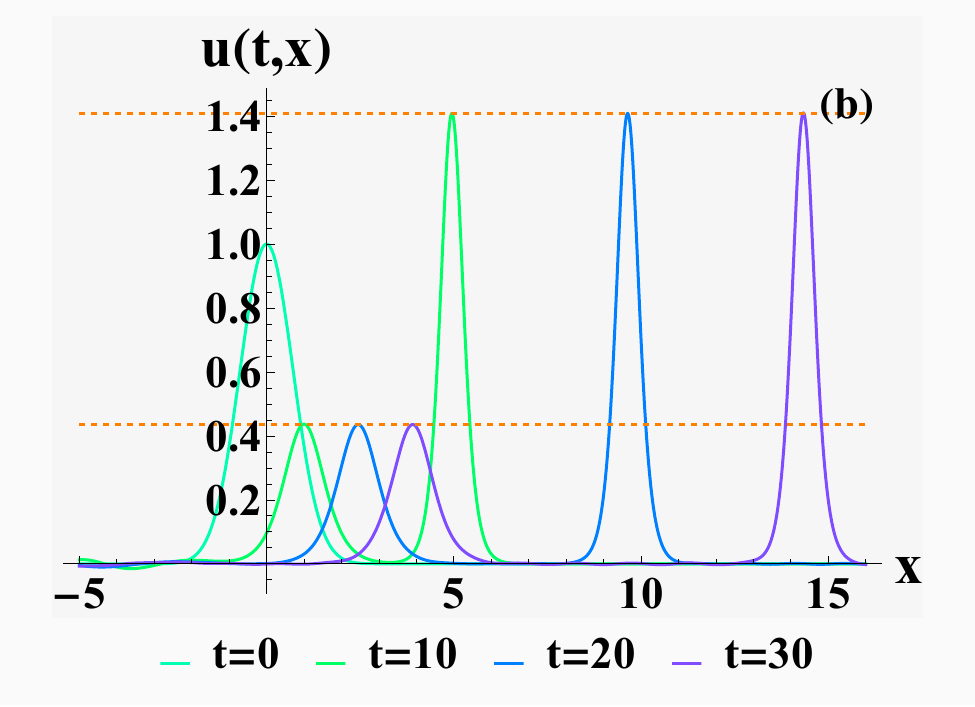}
\includegraphics[scale=0.25]{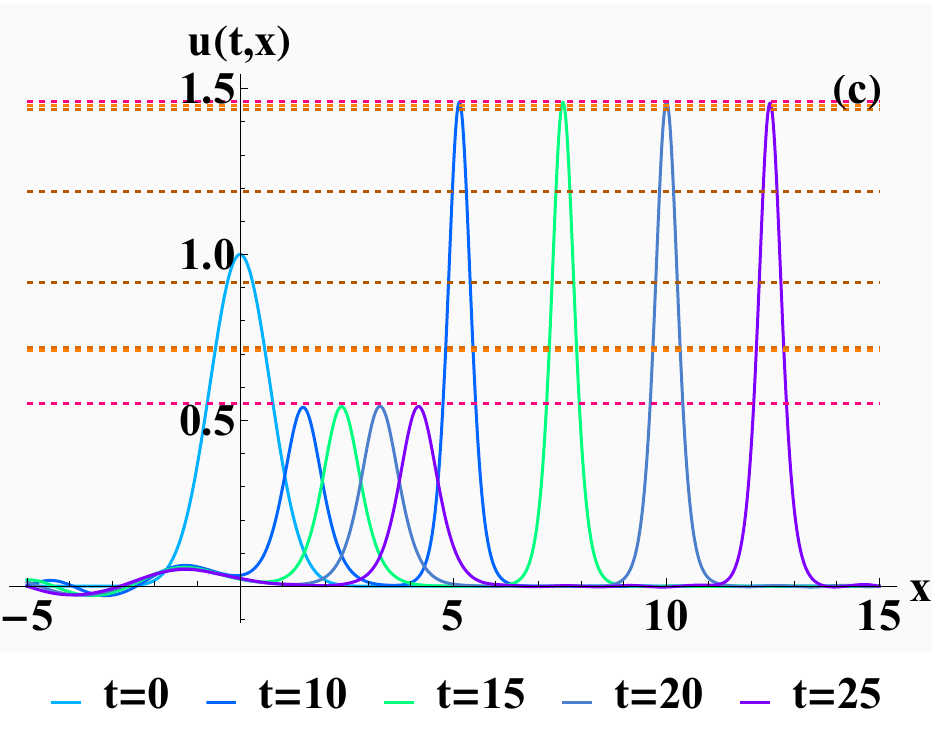}    
\end{minipage}
\caption{In panel (a) predictions of the amplitude for changing values of sigma are plotted for the indicated combinations of conservation laws. The blue, purple, red, orange regions indicate what we find to be the $n=2$, $n=3$, $n=4$ and $n=5$ regions respectively. The non soliton region is shown in green. The $n=4$ and $n=5$ predictions are found by numerically solving the equations, in all the other cases the equations have been solved analytically. The black crosses indicated the amplitudes of emergent solitons from the numerical evolution of a Gaussian initial pulse. In panels (b) and (c) the numerical evolution of a Gaussian initial pulse with  $\sigma=7.22$, $\sigma=8$ respectively are plotted. In panel(b) the orange dotted lines indicate the amplitudes predicted by the relations between $a$ and $\sigma$ plotted in panel (a) in the two soliton region. In panel (c) differently shaded the orange lines indicate the different predictions of the amplitude the same relations make when $\sigma=8$. The pink lines indicate the predictions made by the three soliton region.}
\label{ampsigmakdvfig}
\end{figure}

Thus the argument of \cite{Berezin1967} is that only in this range will a Gaussian initial pulse break up into two solitons. Furthermore the reigon of $\sigma$ for which a Gaussian pulse will break up into $n$ is governed by the conservation laws, and this region can be identified by taking $n$ conservation laws, generating $n$ simultaneous equations and solving for $n$ seperate amplitudes, as done above in the $n=2$ case  \cite{Jeffrey1972}. From these solutions  we should also be able to predict the amplitude of the emergent solitons for a given value of $\sigma$. Numerically evolving a Gaussian initial pulse for a different values of $\sigma$ produces results that are in general agreement with this argument, however the calculations also suggest that some refinements may be neccessary. In \cite{Berezin1967} and \cite{Jeffrey1972} only the lowest possible combinations of conservation laws are considered. However any combination of conservation laws would be equally as valid and given that this system admits an infinite number of conservation laws we have infinitely many combinations available, all generating different relations between $\sigma$ and $a_i$. In figure \ref{ampsigmakdvfig} we plot different predictions of $a_i$ for a given value of $\sigma$, using different combinations of conservation laws. As can be seen in each region apart from the first, the different combinations give roughly the same predictions of the amplitude but they do not agree exactly.  We conjecture that this is the reason for the discrepancy in some cases between the predicted values of the amplitudes, and the amplitudes of the solitons that emerge from numerically evolving a Gaussian initial pulse, which are indicated by the black crosses. We also note that for all regions above $2 \leq n$ there is some encroachment onto the neighbouring region. For instance, there is some overlap between the solution of $n=2$ and $n=3$. Looking at figure \ref{ampsigmakdvfig} we see that the prediticions in these reigons meet at $\sigma=7.22$, where the amplitude of the third soliton in the the three soliton reigon has fallen to zero. There is also at this point a high degree of agreement between the predictions from the different conservation law combinations, hence the accuracy of the predictions (see panel (b) in figure \ref{ampsigmakdvfig}). However moving to higher $\sigma$ we find even if there exists a two soliton prediction of the amplitude it is not at all accurate, whereas the three-soliton prediction is, see for example figure  \ref{ampsigmakdvfig} panel (c) where $\sigma=8$. Indeed it seems that in every reigon $u_0$ breaks up into the maximum number of solitons possible.  No one soliton region exists as the predicted amplitudes are very different until we reach roughly $\sigma=3$ at which point we see the emergence of one soliton but with a significant oscillatory tail.  We find the same behaviour for the mKdV equation, for more details see \cite{stability}.

\begin{figure}[h]
\centering
\begin{minipage}{0.6\textwidth}
\includegraphics[scale=0.62]{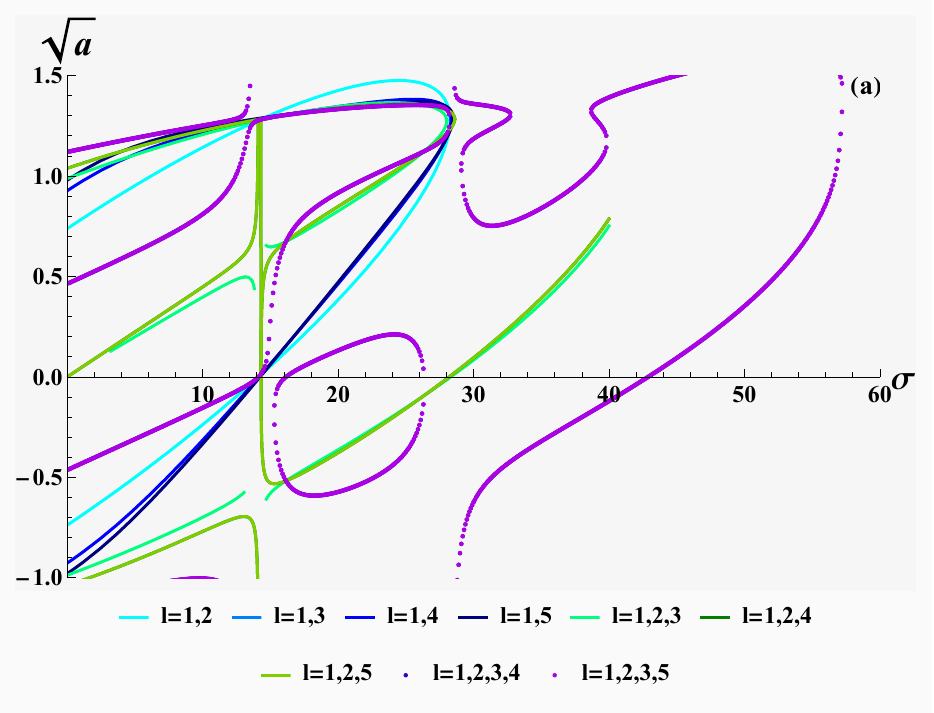}    
\end{minipage}
\begin{minipage}{0.39\textwidth}
\centering
\includegraphics[scale=0.3]{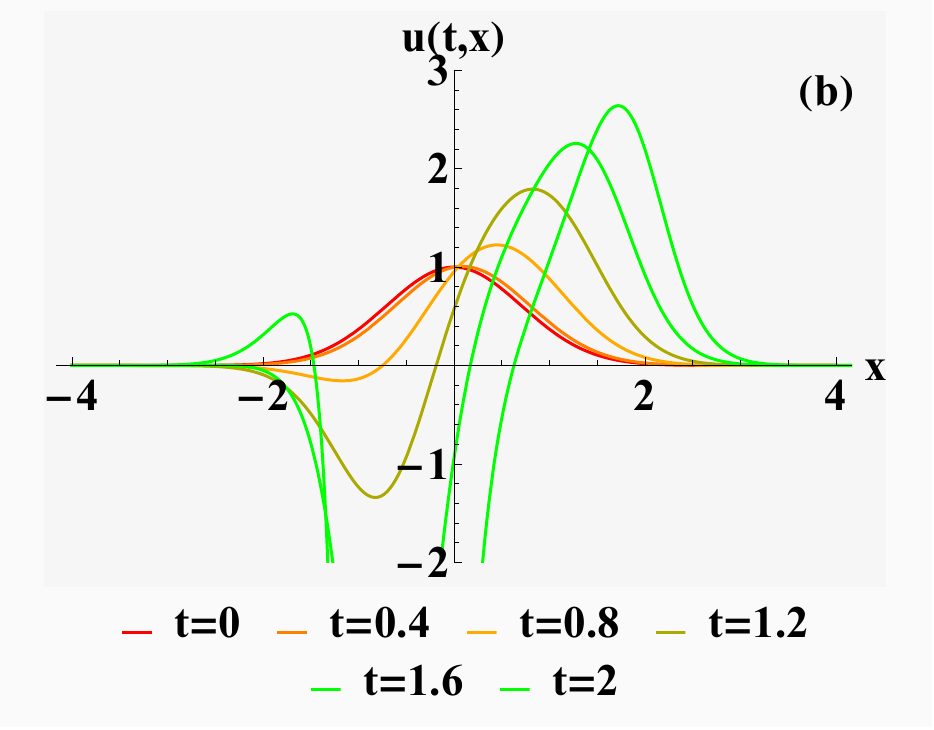}  
\includegraphics[scale=0.3]{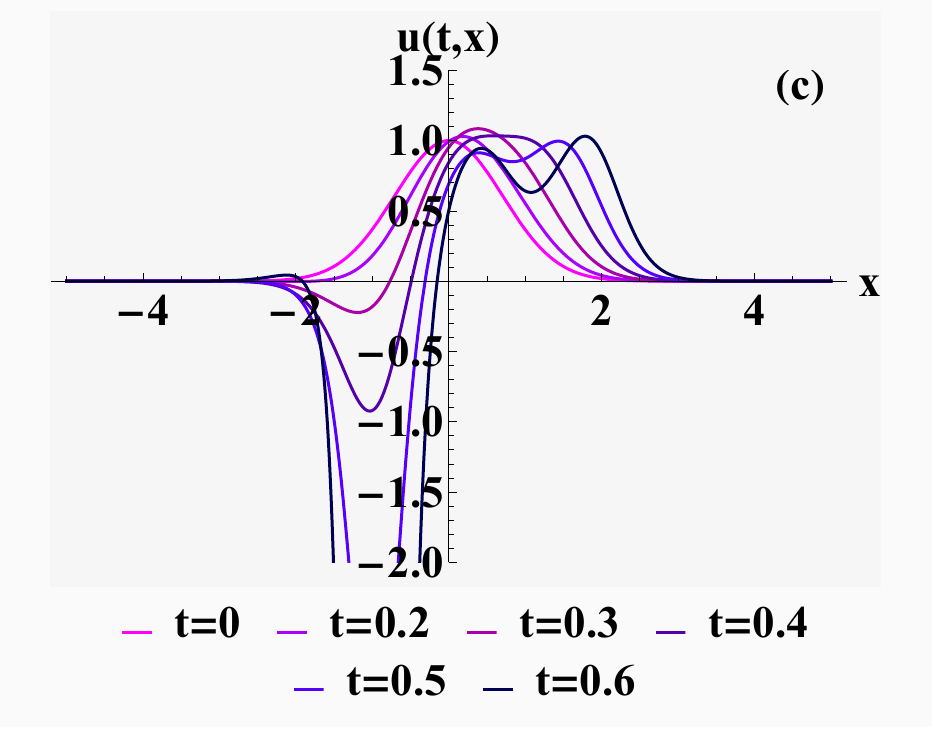}
\end{minipage}
\caption{Panel (a) displays the predictions for the square root of the soliton amplitudes for different combinations of conservation laws when a Gaussian initial pulse is evolved under a rotated KdV equation. In panel (b) and (c) this numerical evolution is displayed for $\sigma=3$ and $\sigma=15$ respectively.}
\label{ampsigmaKdVrot}
\end{figure}

We now want to apply the same argument to the case of the rotated KdV and MKdV equations. Once more we solve (\ref{KdVConservation}) for different combinations of the conservation laws. Note that in the process of interchanging $x$ and $t$ the role of the charge and flux densities are also interchanged, for instance the lowest order conserved charged densities for the rotated KdV equation are given by 

\begin{subequations}
\begin{minipage}{0.48\textwidth}
\begin{equation}
\mathcal{Q}_1^r=\frac{1}{2}u^2+\frac{1}{\sigma^2}u_{2t},
\end{equation}
\end{minipage}
\begin{minipage}{0.48\textwidth}
\begin{equation}
\mathcal{Q}_2^r=\frac{1}{3}u^3+\frac{1}{2\sigma^2}\left(2uu_{2t}-u_{2t}^2\right)	.
\end{equation}
\end{minipage}
\end{subequations}
Repeating the same process as above, we find that only in the "two soliton" region are all amplitudes real and positive. Outside the region given by 
\begin{equation}
\frac{16 \times 2^{\frac{3}{4}} \times 3^{\frac{1}{4}}}{\sqrt{\pi}} < \sigma \leq \frac{32\times 2^{\frac{3}{4}} \times 3^{\frac{1}{4}}}{\sqrt{\pi}},
\label{rangerotated}
\end{equation}
 one or more of the amplitude square roots predicted by the conservation laws are either negative or complex, see figure \ref{ampsigmaKdVrot}. We therefore only expect to see two soliton solutions. However we see that the development of a singularity at the origin kills the evolution of this system before it can evolve into two distinct solitons, which is consistent with what we observe for the exact initial profile discussed above and displayed in figure \ref{kdvanvnum}.
However, we see a different situation for the mKdV equation, once more we see that amplitudes are only real for the range given in (\ref{rangerotated}), for any combination of the conservation laws. However this time we can allow the amplitude of the solitons to be negative, since this is still a solution of the mKdV equation. This means that in the region we can admit negative amplitudes and keep a soliton solution. Therefore what we expect to see in the region in (\ref{rangerotated}) is the initial pulse breaking up into any number of solitons with positive and negative amplitudes. Outside this region we observe the same behaviour, while we have only plotted the four soliton solution, we suspect that other solutions are also valid in this region. We conjecture this is what guards against the kind of divergence we see with the KdV equation, the system does not evolve into a singularity because it is trying to break up into an arbitrary number of solitons.

\begin{figure}[h]
\begin{minipage}{0.6\textwidth}
\includegraphics[scale=0.62]{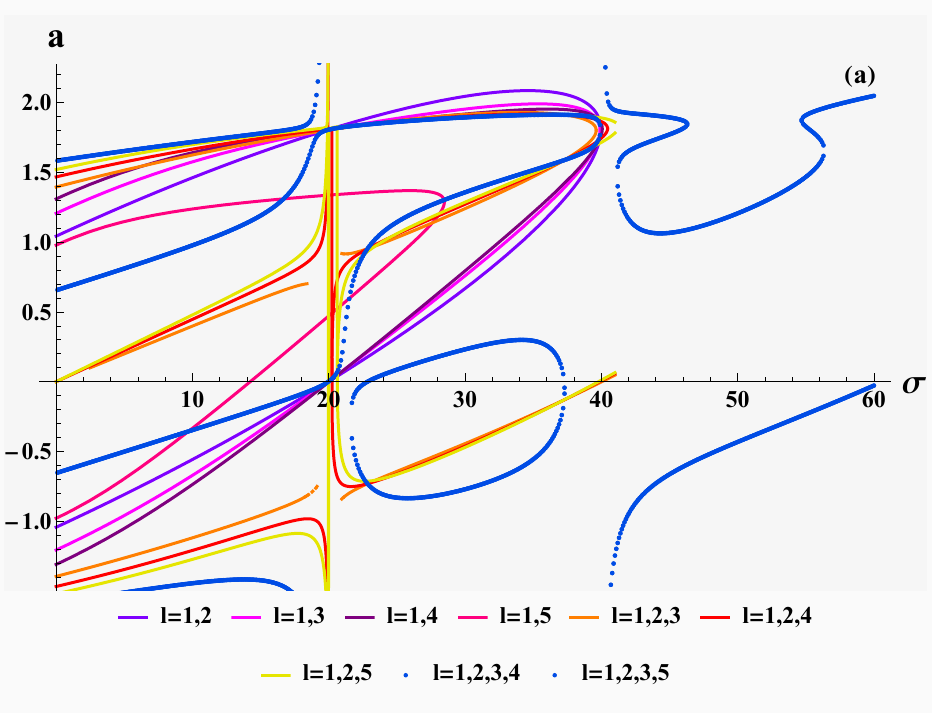} 
\end{minipage}
\begin{minipage}{0.39\textwidth}
\includegraphics[scale=0.35]{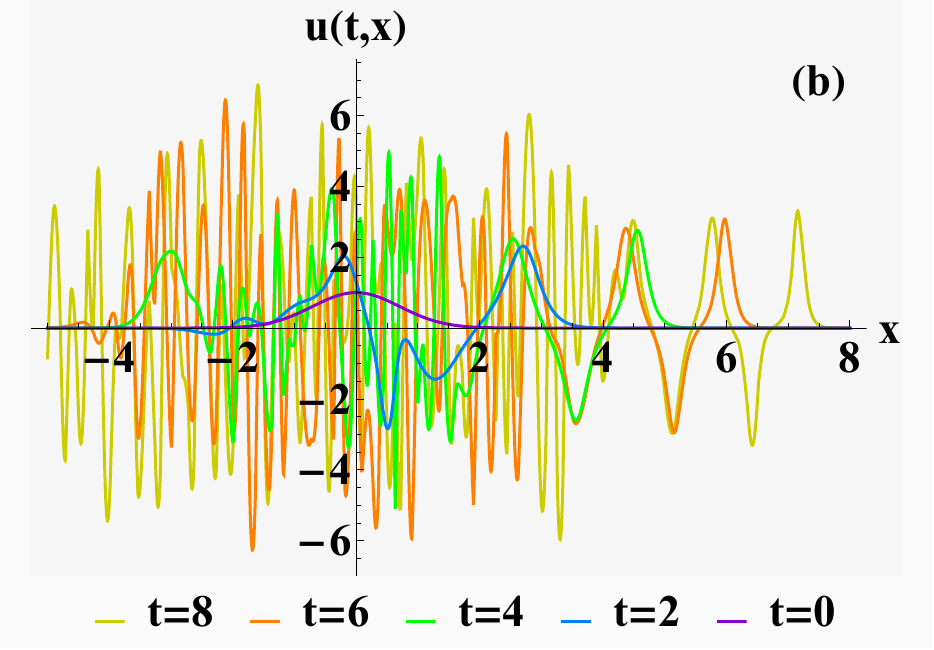}
\includegraphics[scale=0.35]{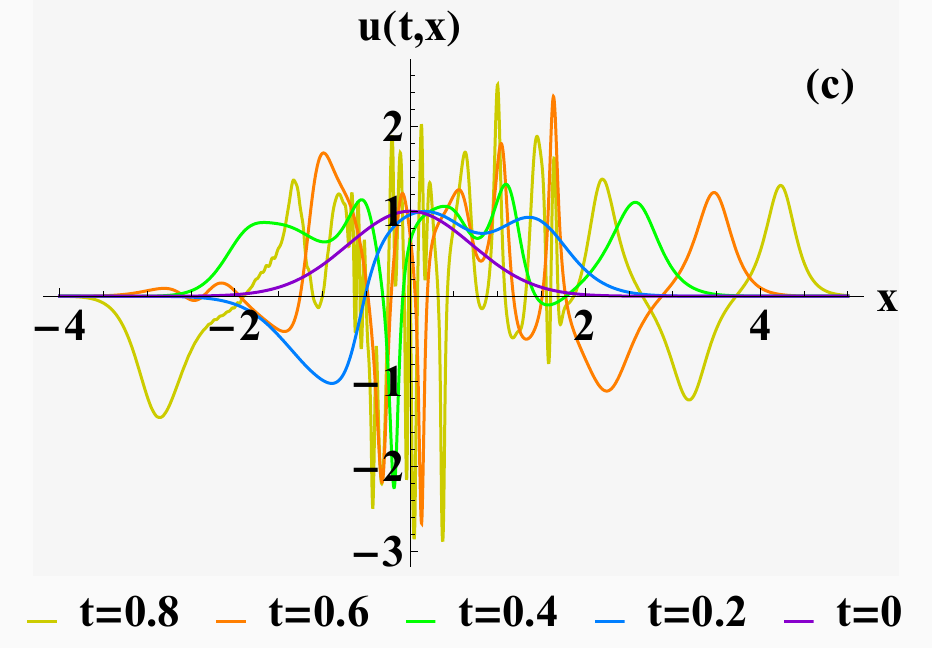}
\end{minipage}
\caption{Panel (a) displays the predictions for the square root of the soliton amplitudes for different combinations of conservation laws when a Gaussian initial pulse is evolved under a rotated mKdV equation. In panel (b) and (c) this numerical evolution is displayed for $\sigma=5$ and $\sigma=45$, respectively.}
\label{ampomegaMKDVrot}
\end{figure}

\section{Quantisation of a linearised Rotated KdV System}
We now quantise a simplified version of the rotated KdV system. The case of $n=2$ is a linear non-integrable system, but we consider it here because its linearity means it can be much more easily quantised, so we treat this as a warm up exercise that will hopefully lead to the quantisation of the $n=3,4$ cases. We follow a procedure for the cannonical quantisation of higher time derivative theories outlined in \cite{Weldon:2003by}. Of course there are many quantisation procedures other than the cannonical but we have not thus far considered them. As a first step in our quantisation we Fourier transform the Lagrangian in (\ref{Lagrangian}) to find the momentum space Lagrangian given by 
\begin{align}
L\left(t,k\right)&= C_0 \psi^2+C_1\psi_t^2+C_2 \psi_{2t}^2, & C_0&=-1, & C_1&=1-\frac{1}{2}\sqrt{1-4k^2}, &C_2=1+\frac{1}{2}\sqrt{1-4k^2}.
\label{LagrangianKSpace}
\end{align}    
From (\ref{LagrangianKSpace}) we then derive the equations of motion

\begin{subequations}
\begin{minipage}[m]{0.32\linewidth}
\begin{align}
0&=C_0\psi+C_1\psi_{2t}+C_2\psi_{4t},
\label{EOMKSpace}
\end{align}    
\end{minipage}
\begin{minipage}[m]{0.32\linewidth}
\begin{align}
0=&C_0-C_1 \omega_l^2 +C_2 \omega_l^4,
\label{QuantEOMOmega}
\end{align}
\end{minipage}
\begin{minipage}[m]{0.32\linewidth}
\begin{align}
\psi=&\sum_{j=1}^2 a_je^{-\omega_j t}+b_je^{\omega_j t}
\label{modeexpansion}
\end{align}
\end{minipage}  
\end{subequations}
where (\ref{EOMKSpace}) is the equation of motion, which can be rewritten as in equation (\ref{QuantEOMOmega}) (note that where $l=1,2$), if we insert the mode expansion in (\ref{modeexpansion}). Thus we have four possible solutions for $\omega$.  We will discuss these in more detail below but first we make some comments on the Hamiltonian and the quantisation. 
Following the same construction as outlined in section 2, we arrive at the Hamiltonian
\begin{align}
H&=-\frac{1}{2}\left(C_0\psi^2+C_1\psi_t^2+2C_2\psi_t\psi_{3t}-C_2\psi_{2t}^2\right), \\
R_i&= \left[2 C_N \omega_i \left(\omega_i^2-\omega_j^2\right)\right]^{-1} \quad i \neq j, \label{QuantumHamiltonian} \\
&=\sum_{i=1}^2\frac{\omega_i}{R_i}b_i a_i + \frac{1}{2}\omega_i. \nonumber
\end{align}

where the rearranging in equation (\ref{QuantumHamiltonian}) is achieved by inserting (\ref{modeexpansion}) and doing some considerable rearranging (see \cite{Weldon:2003by} for details). Already we see in (\ref{QuantumHamiltonian}) that at the classical level the Hamiltonian is not bounded from below, since whether or not we choose $\omega_i$ to have the same or opposite sign the two terms in the summation will have an opposite contribution. Thus we can define one of these modes to be of positive energy and the other to be off negative energy. We now come to the quantisation and as usual we upgrade the Poisson bracket to a commutator, but with the crucial difference that now, as discussed in previous sections, we have additional phase space variables and therefore additional Poisson brackets. Upgrading the coefficients in (\ref{QuantumHamiltonian}) to operators we find that we now have two sets of creation and annihilation operators, corresponding to the two different modes and satisfying the relation
\begin{align}
\left[\hat{a}_m,\hat{b}_n\right]=&-i R_m\delta_{m,n} &
\hat{b}_n=&\hat{a}^\dagger_n & m,n=1,2.
\end{align}

We now come to the key issue when it comes to quantising HTDTs. As we will see we have to choose whether to sacrifice the boundedness of the Hamiltonian or the unitarity of the theory. Which choice we make depends on how we define the groundstate. Say we want to ensure correspondence with the classical case, and keep $\omega_1$ as the positive energy mode and $\omega_2$ as the negative energy mode (vice-versa would also be valid) and thus define the ground state accordingly, we find is that the norms of one particle states and the spectrum of energy eigenstates are given as follows
\begin{align}
\hat{a}_1\ket{0}=\hat{a}_2^\dagger\ket{0}=0 &, & R_1=\bra{0}\hat{a}_1\hat{a}_1^\dagger\ket{0}&, & R_2=\bra{0}\hat{a}_2^\dagger\hat{a}_2\ket{0} &,&  E=n_{r_{1}}\omega_{r_{1}}-n_{r_{2}}\omega_{r_{2}}+\frac{1}{2}\left(\omega_{r_{1}}-\omega_{r_{2}}\right) \nonumber.
\end{align}
Since by equation (\ref{QuantumHamiltonian}), $R_1$ and $R_2$ must be of opposite sign, all norms are positive definite and unitarity is preserved the spectrum of energy eigenstates however is not bounded from below. Conversely we can instead make both energy modes "positive" and thus we have
\begin{align}
\hat{a}_1\ket{0}=\hat{a}_2\ket{0}=0 &, & R_1=\bra{0}\hat{a}_1\hat{a}_1^\dagger\ket{0}&, & R_2=\bra{0}\hat{a}_2\hat{a}_2^\dagger\ket{0} &,&  E=n_{r_{1}}\omega_{r_{1}}+n_{r_{2}}\omega_{r_{2}}+\frac{1}{2}\left(\omega_{r_{1}}+\omega_{r_{2}}\right) \nonumber.
\end{align}
Thus, we have a floor to the energy eigenvalue spectrum, but also states of negative norm, so we have lost unitarity. So whatever choice we make we end up in an undesirable situation. There are some strong opinions on which choice is worse, see \cite{Orstogradskytheorem} for one side and \cite{stelle} for a circumstance where the other choice is preferred, but the key point seems to be that we have a choice about which property we want to keep. 

Returning to our specific problem we solve equation (\ref{QuantEOMOmega}) and find the solutions

\begin{subequations}
\begin{minipage}{0.495\textwidth}
\begin{align}
\omega_1^{\pm}=\pm2\sqrt{\frac{1}{2-\sqrt{4-k^2}-\sqrt{24-k^2+4\sqrt{4-k^2}}}}
\label{omegasol1}
\end{align}
\end{minipage}
\begin{minipage}{0.495\textwidth}
\begin{align}
\omega_2^{\pm}=\pm2\sqrt{\frac{1}{2-\sqrt{4-k^2}+\sqrt{24-k^2+4\sqrt{4-k^2}}}} \nonumber. 
\label{omegasol2}
\end{align}
\end{minipage}
\label{omegasol}
\end{subequations}
Thus we have two choices for how we define the energy eigenvalue spectrum 

\begin{subequations}
\begin{minipage}{0.495\linewidth}
\begin{equation}
E_{n_{1},n_{2}}=n_1\omega_1^{\pm}+n_2\omega_2^{\pm},
\label{ResonanceEnergy1}
\end{equation}
\end{minipage}
\begin{minipage}{0.495\linewidth}
\begin{equation}
E_{n_{1},n_{2}}=n_1\omega_1^{\pm}+n_2\omega_2^{\mp}.  
\label{ResonanceEnergy2}
\end{equation}
\end{minipage}    
\end{subequations}
The first choice in (\ref{ResonanceEnergy1}) combines $\omega$'s
of the same sign and thus the spectrum is bounded, however this will give resonances of opposite sign. The second choice in (\ref{ResonanceEnergy2}) combines $\omega$s of opposite sign and thus the resonances are off the same sign but the spectrum is unbounded. We note that these $\omega$s are everywhere complex. Therefore to try and make some physical sense of this, we examine what would happen if we were to interpret these results as corresponding to resonances (not the same resonance), which in general have energy eigenvalues given by
\begin{align}
E &= E_r - \frac{i}{2}\Gamma,
\end{align}
where $\Gamma$ is the decay width. In order for this to match the results in (\ref{omegasol}) we need the complex parts to be everywhere negative. The only way to achieve this is to choose $\omega_1$ and $\omega_2$ to both be off negative sign, and thus the energy eigenspectrum is the one given in (\ref{ResonanceEnergy1}). Thus we find ourseleves in the situation of having lost unitarity but keeping the boundedness of the spectrum. 

\section{Conclusion}

In the first part we have presented a Hamiltonian formalism of a space-time rotated KdV model. The stability of exact solutions of this system were then studied by comparing their numerical and analytic evolution. The stability of these systems were further explored by considering the evolution of a Gaussian initial pulse.  Based on this study we make the conjecture that the reason or the difference between the evolution of this pulse for the KdV and mKdV equations is that the mKdV equation is symmetric under a change in the overall sign of the solution and the KdV equation is not. In the final section we find that quantising a simplified $n=2$ model brings us to the problem of loosing either unitarity or Hamiltonian boundedness that always arises in HTDT's. It remains to be seen if succesfully quantised $n=3$ and $n=4$ models would exhibit any differences in their quantum behaviour that would in some sense mirror their differences at the classical level.  

\section*{Acknowledgements} The work presented in this article was completed in collaboration with Andreas Fring and Takanoa Taira. BT would like to thank the organisers of ISQS-28 for the opportunity to present at the conference. BT is supported by a City, University of London Research Fellowship. 

\printbibliography

\end{document}